\documentclass[aps, superscriptaddress ,twocolumn,prl]{revtex4}
\usepackage{amsmath}
\usepackage{amssymb}
\usepackage{epsfig}
\usepackage{graphics}
\usepackage{bm}
\usepackage{braket}

\begin{document}
\title{Dipolar fermions in a two-dimensional lattice at non-zero temperature}
\author{Anne-Louise\ Gadsb\o lle}
\affiliation{Lundbeck Foundation Theoretical Center for Quantum System Research}
\affiliation{Department of Physics and Astronomy, University of Aarhus, Ny Munkegade, DK-8000 Aarhus C, Denmark}
\author{G.\ M.\ Bruun}
\affiliation{Department of Physics and Astronomy, University of Aarhus, Ny Munkegade, DK-8000 Aarhus C, Denmark}

\begin{abstract}
We examine density ordered and superfluid phases of fermionic dipoles in a two-dimensional square lattice at non-zero temperature.
 The critical temperature of the density ordered phases is determined and is shown
  to be proportional to the coupling strength for strong coupling. We calculate the superfluid fraction and 
demonstrate that the Berezinskii-Kosterlitz-Thouless transition temperature of the superfluid phase is proportional to the hopping matrix element in the 
strong coupling limit. We finally analyze the effects of an external harmonic trapping potential.  
\end{abstract}
\maketitle

\section{Introduction} 
An increasing number of experimental groups are trapping and cooling atoms or molecules with a permanent magnetic or electric dipole moment. 
Bose-Einstein condensates of $^{52}$Cr atoms~\cite{Lahaye,Koch} and of  
$^{164}$Dy atoms~\cite{Lu} with large magnetic dipole moments have been realized.
Fermionic gases of $^{40}$K$^{87}$Rb~\cite{KRb} and $^{23}$Na$^6$Li~\cite{NaLi} molecules with an electric dipole moment 
have  been created, and  the first steps toward the formation of fermionic
 $^{23}$Na$^{40}$K molecules have been reported~\cite{NaK}. Also,  
 experimental progress toward realizing dipolar molecules in an optical lattice have recently been presented~\cite{Danzl}.
The anisotropy of the dipole interaction results in many intriguing effects. 
In a two-dimensional (2D) lattice, the existence of density ordered phases with a complicated 
unit cell~\cite{Mikelsons}, liquid crystal phases~\cite{Lin}, and a supersolid 
phase~\cite{He} have been predicted when the dipole moments are perpendicular to the lattice plane. Tilting the 
dipoles toward the lattice plane leads to density order with different symmetry,  superfluidity and bond-solid order at zero 
temperature~\cite{Gadsbolle,Bhongale,Danshita}.
When a trapping potential is present, these phases were shown to coexist, forming ring and island structures~\cite{Gadsbolle}.

In this paper, we examine fermionic dipoles in a 2D square lattice including the presence of a harmonic trapping potential. Focus is on the 
effects of a non-zero temperature and the melting of density ordered and superfluid phases. We determine the
critical temperature for the density ordered phases and  find that it  is proportional to the interaction strength in the strong coupling regime.
 For the superfluid phase, we calculate 
the superfluid fraction and the Berezinskii-Kosterlitz-Thouless (BKT) transition temperature, which is proportional to the 
hopping matrix element in the strong coupling limit. 
 We analyze the effects of an external trapping potential showing that for experimentally realistic systems,
  the ordered phases exist in the center of
the trap with melting temperatures close to that which can be obtained from a local density approximation.

\begin{figure}[bh]
\includegraphics[width=0.7\columnwidth]{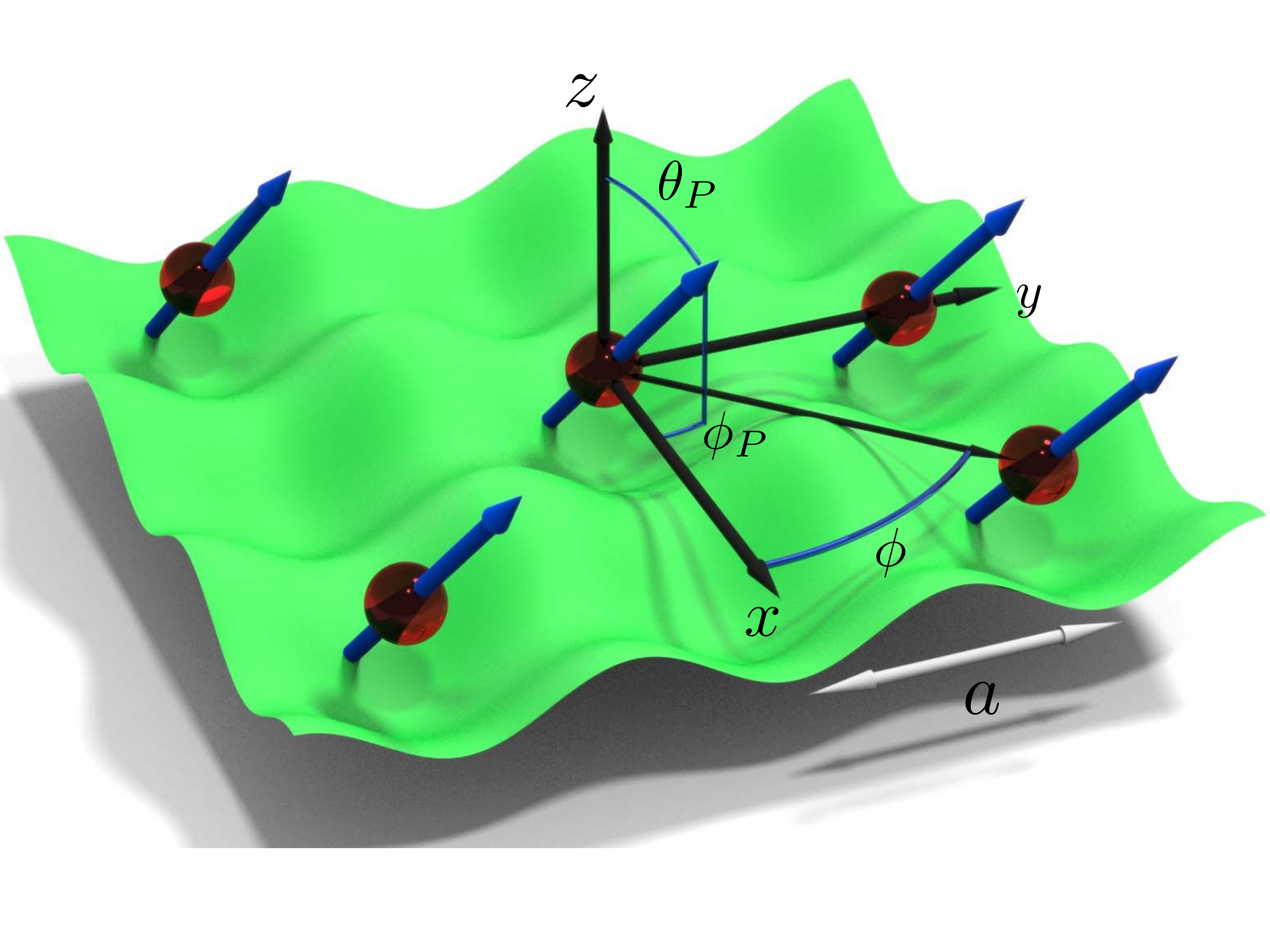}
\caption{(color on-line) Dipoles move in a 2D square  lattice with lattice constant $a$. They are aligned forming an angle $\theta_P$ with the $z$ axis perpendicular to the lattice 
plane, and the azimuthal angle $\phi_P$ with the $x$ axis which is parallel to a lattice vector. }
\label{setup}
\end{figure}
 
\section{Model}
We consider fermionic dipoles of mass $m$ and dipole moment ${\mathbf d}$ moving in a 2D square lattice with lattice constant $a$.
 The dipole moment is aligned by an external field to form an angle $\theta_P$ with respect to the $z$ axis which is perpendicular to the lattice plane and an angle $\phi_P$ with respect to a lattice vector chosen as the $x$ axis. The Hamiltonian is
 $\hat{H}=\hat{H}_{\rm kin}+\hat{V}$ where 
\begin{equation}
\hat{H}_{\rm kin}=-t\sum_{\langle ij\rangle}\left(\hat{c}_{i}^{\dagger}\hat{c}_{j }+h.c.\right)
+\sum_{i}\left(\frac 1 2 m\omega^2r_i^2-\mu\right)\hat{n}_{i}
\end{equation}
and 
 \begin{equation}
\hat{V}=\frac 1 2 \sum_{i\neq j }V_D({\mathbf r}_{ij})\hat{n}_{i}\hat{n}_{j}
\label{Hint}
\end{equation}
where ${\mathbf r}_i$ denotes the position of lattice site $i$ and ${\mathbf r}_{ij}={\mathbf r}_{i}-{\mathbf r}_{j}$,
$\hat{c}_{i}$ is the annihilation operator that removes a dipole at site $i$, and $\hat{n}_{i}=\hat{c}_{i}^{\dagger}\hat{c}_{i}$ is the number operator. 
The chemical potential is $\mu$ and $t$ is the hopping matrix element between nearest neighbors $\braket{ij}$. We include the 
effects of a harmonic potential with trapping frequency $\omega$ exactly in our analysis. The 
interaction between two dipoles separated by ${\mathbf{r}}$ is given by
\begin{gather}
V_D({\mathbf{r}})=\frac{D^2}{r^3}\left(1-3\cos^{2}\theta_{\rm rd}\right)\nonumber\\
=\frac{D^2}{r^3}\left[1-3\cos^{2}(\phi_{P}-\phi)\sin^{2}\theta_{P}\right] 
\label{Interaction}
\end{gather}
with $D^2=d^2/4\pi\epsilon_0$ for electric dipoles and $\theta_{\rm rd}$ the angle between ${\mathbf d}$ and
${\mathbf r}=r(\cos \phi, \sin \phi,0)$, see Fig.\ \ref{setup}. We define $g=D^2/a^3$ as a measure of the interaction strength. 

The anisotropy of the dipolar interaction (\ref{Interaction}) with  attractive and repulsive regions gives rise to both density ordered and superfluid 
phases~\cite{Gadsbolle,Bhongale,Mikelsons}.  We capture the existence of these competing phases
 using mean-field theory including the Hartree terms and the pairing terms, which we expect to be reasonably accurate due to the long range nature of the interaction. 
 The mean-field Hamiltonian is diagonalized by solving the Bogoliubov-de Gennes equations~\cite{Gadsbolle}
\begin{align}
\sum_j\begin{pmatrix}
L_{ij} & \Delta_{ij} \\
\Delta_{ji}^* & -L_{ij}
\end{pmatrix}
\begin{pmatrix}
u_{ \eta }^{j} \\
v_{ \eta}^{j}
\end{pmatrix}
=E_{\eta }
\begin{pmatrix}
u_{ \eta }^{i} \\
v_{\eta }^{i}
\end{pmatrix},
\label{BdGeqn}
\end{align}
where $\Delta_{ij}=V_D({\mathbf r}_{ij})\langle \hat c_{j}\hat c_{i}\rangle$ and 
\begin{gather}
L_{ij}=-t \delta_{\langle ij\rangle}+(\sum_{k}V_D({\mathbf r}_{ik})\langle n_{k}\rangle+\frac m 2 \omega^2r_i^2-\mu) \delta_{ij}.
\end{gather}
Here $\delta_{ij}$ and $\delta_{\langle ij\rangle}$ are
the Kronecker delta functions connecting on-site and nearest neighbor sites, respectively.
Self-consistency is obtained iteratively through the usual relations 
$\langle\hat{n}_{i}\rangle=\sum_{E_\eta>0}\left[(1-f_\eta)|v_{\eta}^{i}|^{2}+f_{\eta}|u_{\eta}^{i}|^{2}\right]$ and
 $\langle\hat{c}_{i }\hat{c}_{j }\rangle =\sum_{E_\eta>0}[u_{\eta}^{i}v_{\eta}^{j*}(1-f_\eta)+v_{\eta}^{i*}u_{\eta}^{j}f_\eta]$, 
 with  $f_\eta=[\exp(E_\eta/T)+1]^{-1}$ the Fermi function for the temperature $T$. We use units where $k_B=\hbar=1$.
 To analyze the melting of the  superfluid phase, we shall use the framework of BKT theory.

\section{Stripe melting at half filling} 
We first analyze the case of no trapping potential and  half filling, $N/N_L=1/2$, with $N_L$ the  number of lattice sites
and $N=\sum_i\langle\hat n_i\rangle$ the total number of particles.
When the dipoles are perpendicular to the lattice, it follows from the perfect nesting of the Fermi surface that a phase with checkerboard density order 
persists down to $g/t\rightarrow 0$ for $T=0$~\cite{Mikelsons}. In the limit of strong interaction  $g/t\gg 1$ where the kinetic energy can be 
neglected and the problem becomes classical, it was shown that the checkerboard phase is replaced by a striped phase when the dipoles are 
tilted at a sufficiently large angle $\theta_p$~\cite{Gadsbolle}. We now examine the melting of these density ordered phases at a 
non-zero temperature. The melting is in the Ising universality class due to the discreteness of the lattice, and we therefore expect mean-field theory to yield 
a qualitatively correct value for the transition temperature.

For the case of stripes along the $x$ direction, we express the density as  $\braket{\hat{n}_{i}}=1/2\left[1+M(-1)^{y_{i}/a}\right]$ with 
$M$ the order parameter. The corresponding mean-field Hamiltonian can be written as 
$
\hat H=\sum_{k_y>0}\left[E_{1{\mathbf k}}\gamma^\dagger_{1\mathbf{k}}\gamma_{1\mathbf{k}}+
E_{2{\mathbf k}}\gamma^\dagger_{2\mathbf{k}}\gamma_{2\mathbf{k}}\right]
$
with the single particle energies 
\begin{align}
E_{1 \mathbf{k} }=\xi_{\mathbf k}+ \sqrt{(2t\cos k_{y}a)^2+ [\tilde V_D(0,\pi/a)M/2 ] ^{2}} 
\label{QPenergy}
\end{align} 
where $\xi_{\mathbf k}=-2t\cos k_{x}a -\mu+\tilde V_D(0,0)/2$.  
We have defined the Fourier transform 
$\tilde V_D({\mathbf k})=\sum_i\exp(-i{\mathbf k}\cdot{\mathbf r}_i)V_D({\mathbf r}_i)$.
 The energy $E_{2{\mathbf k}}$ is given by (\ref{QPenergy}) with a minus-sign in front of the square root. 
The self-consistency equation reads 
\begin{align}
1 =\frac{1}{N_{L}}\sum_{k_{y}>0}\frac{\tilde V_D(0,\pi/a)(f_{1\mathbf{k}}-f_{2\mathbf{k}} ) }{ \sqrt{(2t\cos k_{y})^2+ [\tilde V_D(0,\pi/a)M/2] ^{2}}}.
\label{SELFCONS}
\end{align}
where the sum is over half the first Brillouin zone with $k_y>0$.
 In the limit of strong interaction $g/t\gg 1$, Eq.\ (\ref{SELFCONS}) yields 
 \begin{align}
T_{c}^{\rm st}=-\frac1 4 \tilde V_D(0,\pi/a).
\label{TcSt}
\end{align}
When the dipoles are aligned in the lattice plane with $(\theta_p,\phi_p)=(\pi/2,0)$, Eq.\ (\ref{TcSt}) gives
 $T_{c}^{\rm st} \approx 1.27g$. 
A similar analysis for the checkerboard phase yields $T_{c}^{\rm cb}=-\tilde V_{D}(\pi/a,\pi/a)/4$ in the strong coupling 
limit, which gives $T_{c}^{\rm cb} \approx 0.66  g$ for  $\theta_p=0$~\cite{Mikelsons}.

\begin{figure}[th]
\includegraphics[clip=true,width=1\columnwidth]{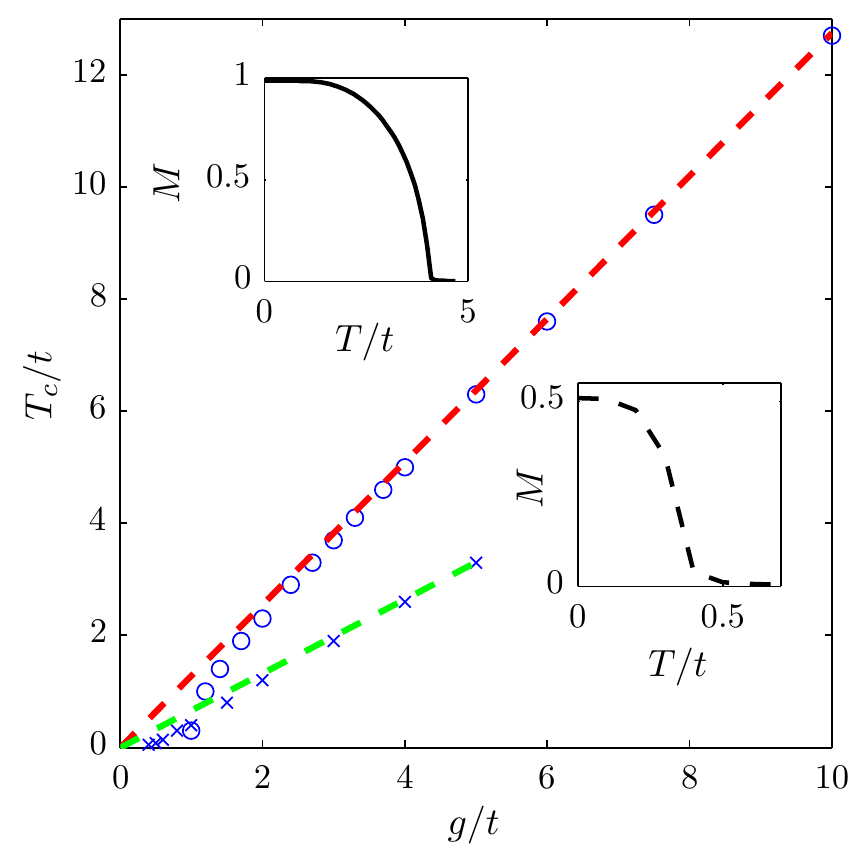}
\caption{(color on-line) The critical temperature of the striped phase for $(\theta_P,\phi_P)=(\pi/2,0)$ ($\circ$'s) and of the checkerboard phase 
 for $\theta_P=0$  ($\times$'s) as a function of coupling strength for half filling obtained from a numerical calculation on $30 \times 30$ lattice sites.
 The dashed lines give the strong coupling results $T_{c}^{\rm st}=-\tilde V_D(0,\pi/a)/4$ with $(\theta_{P},\phi_{P})=(\pi/2,0)$ and  $T_{c}^{\rm cb}=-\tilde V_{D}(\pi/a,\pi/a)/4$ with $\theta_{P}=0$.
 The upper left inset shows the striped order parameter $M$ for $g/t=3.3$ as a function of $T$ for $(\theta_{P},\phi_{P})=(\pi/2,0)$. The lower right inset shows the 
 checkerboard order parameter $M$ for $g/t=1$ as a function of $T$ for $\theta_P=0$. }
 \label{Tc-DW-CB}
\end{figure}

\begin{figure}[th]
\includegraphics[clip=true,width=1\columnwidth]{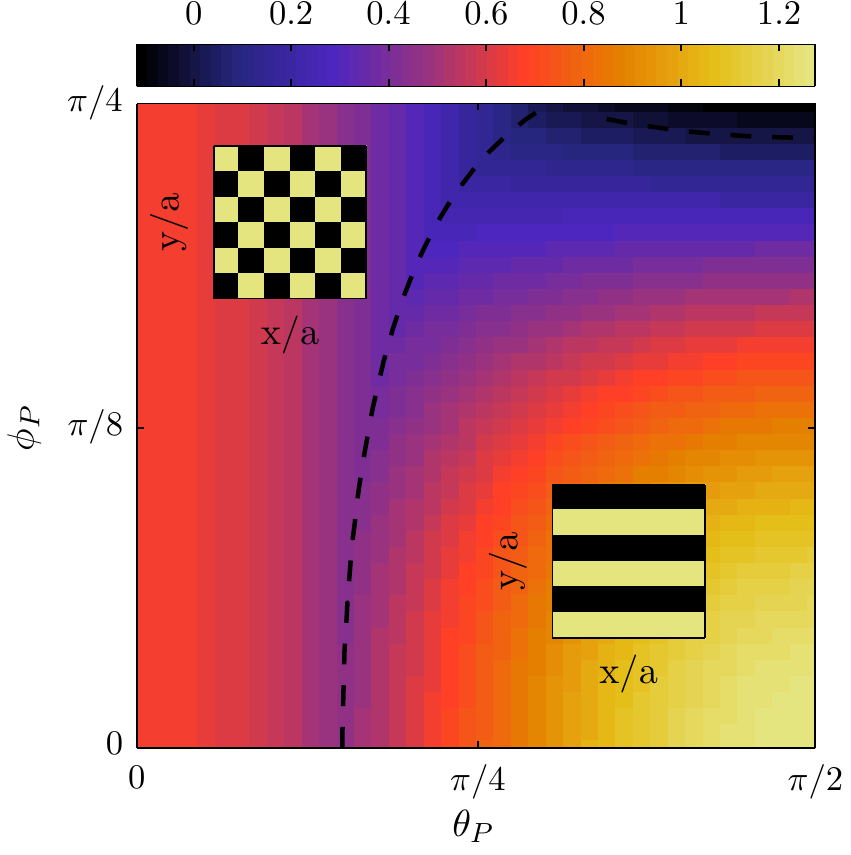}
\caption{(color on-line) The critical temperature in units og $g$ of the striped and checkerboard phases as a function 
of the dipole orientation $(\theta_P,\phi_P)$ for half filling. A dashed line marks the boundary between the stripe and checkerboard phases, and 
the region with no density order is bounded by another dashed line. }
 \label{Tc-phasediagram}
\end{figure}

Figure \ref{Tc-DW-CB} shows the critical temperature as a function of the interaction strength for the checkerboard phase with $\theta_P=0$ 
and for the striped phase with $(\theta_P,\phi_{P})=(\pi/2,0)$. The $\circ$'s and $\times$'s 
are numerical results for the stripe and checkerboard phases respectively, obtained from solving (\ref{BdGeqn}),
and the lines the analytical results for the strong coupling limit discussed above. 
 Finite size effects of the system are eliminated  by neglecting the high temperature tail of the order parameter.
 For example, for the lower right inset in Fig. \ref{Tc-DW-CB} the elimination of the high temperature tail  gives the critical temperature $T_c^{\rm cb}/t=0.4$. 
We see that the numerical results agree well with the strong coupling results for $g/t\gg 1$ whereas the critical temperature becomes exponentially suppressed 
in the weak coupling limit. Note that the critical temperature of the striped phase is almost twice that of the checkerboard phase, which makes it easier to 
observe experimentally.  The upper left inset 
shows how the striped order parameter $M$ decreases with $T$ for $(\theta_P,\phi_{P})=(\pi/2,0)$ and $g/t=3.3$, and the lower right inset shows 
the checkerboard order parameter $M$ as a function of $T$ for $\theta_P=0$ and $g/t=1$.

Figure \ref{Tc-phasediagram} shows the critical temperature of the striped and the checkerboard phase as a function 
of $(\theta_P,\phi_P)$ in the strong coupling regime. It is obtained from   
$\max[-\tilde V(0,\pi/a)/4,-\tilde V(\pi/a,\pi/a)/4]$. For  most orientations of the dipoles, the 
critical temperature of the striped phase exceeds that of the checkerboard phase. We note that the 
upper left corner in the phase-diagram shows a negative critical temperature which indicates that none of the two phases we explore are stable in this region.

\section{Stripe and superfluid melting at one third filling}
For smaller filling fractions, the system can be in a superfluid state  with $p$-wave symmetry for large enough $\theta_P$~\cite{Gadsbolle,Bhongale}.
This leads to a competition between density and superfluid order in analogy with dipoles moving in a 2D plane without a
 lattice~\cite{BruunTaylor,Stripes}.
As an example, we now 
consider the melting of the superfluid and the striped phase for the 
filling fraction $N/N_L=1/3$ and $(\theta_P,\phi_P)=(\pi/2,0)$. For these parameters, mean-field theory predicts the system to be 
superfluid for $g/t\le1.15$ and to exhibit stripe order for $g/t>1.15$ at $T=0$~\cite{Gadsbolle}.

For the 2D system considered here, the melting of the superfluid phase is of the 
BKT type with a transition temperature determined by the phase stiffness of the order parameter~\cite{KosterlitzThouless,Chaikin}.
The phase stiffness $J_x$ associated with a phase twist of the superfluid order parameter in the $x$ direction is determined from the energy 
cost 
\begin{equation}
F_\Theta-F_0 \simeq \frac {J_x} 2\sum_{i}\delta\Theta^2.
\label{FreeEnergy}
\end{equation}
 Here, $F_\Theta$ is the free energy when the phase of the order parameter varies by $\delta\Theta$
between neighboring sites in the $x$ direction and $F_0$ is the free energy when there is no phase twist~\cite{Fisher}. 
Associated with the phase twist,  we 
 define the superfluid fraction $\rho_{s,x}$ by writing 
\begin{align}
F_{\Theta}-F_{0} = \frac{N}{2}\rho_{s,x}m^{*}{v_s}^2=\frac{N}{4} t \rho_{s,x}\delta \Theta^2,
\label{fs}
\end{align}
where $v_{s}=\delta\Theta/2m^*a$ is  the superfluid velocity of the Cooper pairs with mass $2m^*$. The effective mass 
for the dispersion $-2t(\cos k_{x}a +\cos k_{y}a)$ is 
$m^*=1/2ta^2$. Note that the superfluid fraction is dimensionless. 
Similar expressions hold for the phase stiffness $J_y$ and the  superfluid fraction $\rho_{s,y}$ for the $y$ direction.

A linear phase twist along the $x$ direction is equivalent to 
acting on the Hamiltonian with the unitary gauge transformation 
\begin{equation}
\hat H_\Theta=e^{-i\delta\theta \sum_l \hat x_l/a}
 \hat H 
 e^{i\delta\theta \sum_l \hat x_l/a}
 \label{Hgauge}
\end{equation}
  where $x_l$ is the $x$-coordinate of 
particle $l$~\cite{Lieb}. We have $\delta\Theta=2\delta\theta$
since  the superfluid order parameter  involves two particles so that the gauge transformation gives 
 $\Delta_{ij}\rightarrow \Delta_{ij}\exp[{i(x_i+x_j) \delta\theta/a}]$. The gauge transformation only affects  
 $\hat H_{\rm kin}$ by introducing
 a phase factor $t\hat{c}_{ i}^{\dagger}\hat{c}_{i\pm e_x}\rightarrow te^{\pm i  \delta \theta}\hat{c}_{ i}^{\dagger}\hat{c}_{i\pm e_x}$
 on the hopping terms connecting neighboring sites in the $x$ direction. Here,  $e_x$ denotes one lattice step in the $x$ direction.
  Since we only need the energy cost to lowest order in the 
 phase twist to determine $J$ from Eq.\ (\ref{FreeEnergy}), it is sufficient to  use perturbation theory in $\delta\theta$.
 Expanding to second order in $\delta \theta$, we obtain $\hat H_\Theta=\hat H+\hat J+\hat T$ with 
 \begin{eqnarray}
\hat J=-i\delta\theta t\sum_{  i} \left(\hat{c}_{ i }^{\dagger}\hat{c}_{ i+e_x}-\hat{c}_{ i}^{\dagger}\hat{c}_{i-e_x}\right) \nonumber\\
\hat{T} =\frac t 2 \delta\theta^2\sum_{i}\left(\hat{c}_{ i}^{\dagger}\hat{c}_{ i+e_x}+\hat{c}_{ i}^{\dagger}\hat{c}_{ i-e_x}\right). 
\label{TJ}
\end{eqnarray} 
 Since the unitary transformation conserves particle number, we can take  $F_\Theta-F_0=\Omega_\Theta-\Omega_0$ 
where $\Omega=F-\mu N$ with $N$ the total number of particles~\cite{Taylor}. The linked cluster expansion  gives~\cite{Mahan}
   \begin{equation}
\Omega_\Theta-\Omega_0 =\braket{\hat T}-\frac \beta 2 \braket{\hat J^2}
\label{LinkedCluster}
\end{equation}
where $\braket\ldots$ denotes the thermal average with respect to the untwisted Hamiltonian and we have used that there is no current in the 
untwisted case, i.e. $\braket{\hat J}=0$. Mean-field theory gives after some lengthy but straightforward algebra 
\begin{equation}
\braket {\hat{T}}=\frac t 2 \delta\theta^2\sum_{\eta, i }(u_{\eta} ^{i*}u_{\eta }^{i+e_x}+u_{\eta }^{i*}u_{\eta}^{i-e_x})f_{\eta}
\label{TMF}
\end{equation}
and
\begin{gather}
\left\langle J^{2}\right\rangle =-t^{2} \delta\theta^2 \sum_{ij} 
\sum_{\eta \alpha} \sum_{k,l=-1}^{1}kl \left[u_{\eta }^{i*}u_{\alpha }^{j*} u_{\eta }^{j+ke_x}u_{\alpha }^{i+le_x}\right.\nonumber\\
 \times\left. f_\eta(1-f_{\alpha})- u_{\eta }^{i*}v_{\eta}^{j}u_{\alpha }^{i+ke_x}v_{\alpha}^{j+le_x *} f_{\eta}(1-f_{\alpha})\right]. 
 \label{JMF}
\end{gather}
The  sums in Eqs. (\ref{TMF})-(\ref{JMF})
are  taken over positive as well as negative energies, and we have made use of the duality 
$(u_\eta,v_\eta, E_\eta) \leftrightarrow (v^*_\eta,u^*_\eta, -E_\eta)$ of the Bogoliubov-de Gennes equations.

When there is no trap, the Bogoliubov-de Gennes equations are straightforward to solve and Eqs.\ (\ref{fs}), (\ref{TMF}), and (\ref{JMF})
yield 
\begin{equation}
\rho_{s,x}=\frac 1 N\sum_{\mathbf k}\left[n_{\mathbf{k}}\cos k_xa-\frac{2t}{T}f_{\mathbf k}(1-f_{\mathbf k})\sin^2 k_xa\right].
\end{equation}
Here $E_{\mathbf k}$ are the  BCS quasiparticle energies for the $p$-wave  paired state,
and $n_{\mathbf{k}}=u_{\mathbf k}^2f_{\mathbf k}+v_{\mathbf k}^2(1-f_{\mathbf k})$. 
In the continuum limit $a\rightarrow 0$ keeping the density $N/N_La^2$ constant, this  reduces to the usual expression
$\rho_{s,x}=1+(3m^*n)^{-1}(2\pi)^{-3}\int d^3k\partial_Ef_{\mathbf k} k^2$ for a  single component superfluid~\cite{Landau}.

From the phase stiffness, we can extract the  transition temperature as 
 $T_{\rm BKT}=\pi \bar J/2$~\cite{KosterlitzThouless,Chaikin} where we have taken the average $\bar J=(J_x+J_y)/2$ to account for the 
anisotropy of the $p$-wave pairing. Equations (\ref{FreeEnergy})-(\ref{fs}) give 
$\bar J=N\bar\rho_s t/2N_L$ with $\bar \rho_s=(\rho_{s,x}+\rho_{s,y})/2$, 
and we finally obtain 
\begin{equation}
T_{\rm BKT}=\frac{\pi}{4}\frac N{N_L}\bar\rho_st=\frac \pi 8 \frac{\bar n_s}{m^*}
\label{TBKT}
\end{equation}
with the superfluid density defined as $\bar n_s=N\bar \rho_s/N_La^2$. 

\begin{figure}[th]
\includegraphics[clip=true,width=0.9\columnwidth]{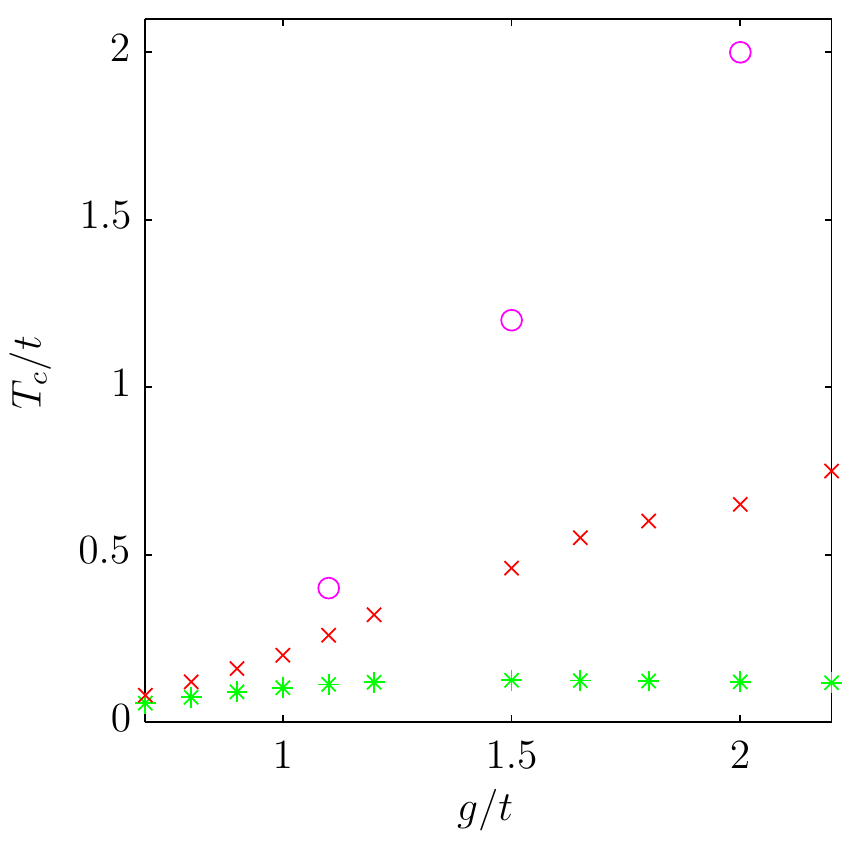}
\caption{(color on-line) The critical temperature  for the superfluid phase  ($*$'s)
 and the striped phase ($\circ$'s) for $(\theta_P,\phi_P)=(\pi/2,0)$ as a function of coupling strength
 for one third filling obtained from a numerical calculation on a $27 \times 27$ lattice site. The  $\times$'s
 give the mean-field superfluid transition  temperature $T^*$. For illustrative purposes, we plot the critical temperature of the superfluid phase even for $g/t>1.15$, 
 where stripe order suppresses superfluidity. 
  }
\label{TcFig}
\end{figure}

In Fig.~\ref{TcFig}, we plot $T_{\rm BKT}$ as a function of the coupling strength obtained from Eq.~(\ref{TBKT}).
 For comparison, we  plot the mean-field superfluid transition temperature $T^*$.
 We also plot the critical temperature $T_c^{\rm st}$ for the stripe phase which is the ground state for $g/t>1.15$.
For weak coupling, the $T_{\rm BKT}$ approaches $T^*$ as expected~\cite{Miyake}, 
whereas it is significantly lower for  stronger coupling. For strong coupling, it follows from Eq.~(\ref{TBKT}) that the critical temperature 
will saturate at $T_{\rm BKT}\sim t$. Indeed, the numerical results yield $T_{\rm BKT}\simeq 0.12t$  for $g/t\gg 1$ as can 
be seen from Fig.~\ref{TcFig}. Note however that stripe order sets in for $g/t> 1.15$ which suppresses the superfluid order. Like 
the case for half filling, we have $T_c^{\rm st}\sim g$ for the critical temperature for the striped phase, which  is a  higher temperature 
than the superfluid transition temperature. It is interesting that  both critical temperatures, $T_{\rm BKT}\sim t$ and $T_c^{\rm st}\sim g$, can be much higher 
than that of the antiferromagnetic phase for atoms in a 3D lattice, which scales as $T_N\sim t^2/U$ in the strong coupling limit 
with $U\gg t$ the on-site interaction~\cite{Duan,TNeel}.

\begin{figure}[th]
\includegraphics[clip=true,width=0.9\columnwidth]{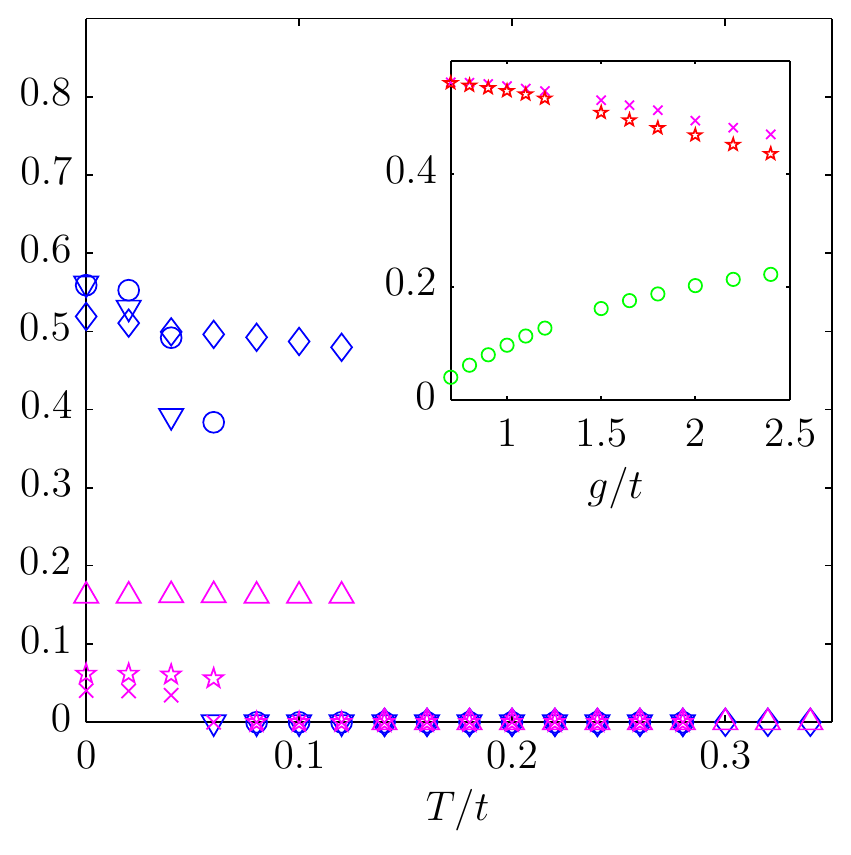}
\caption{(color on-line) The superfluid fraction $\bar\rho_s$ and the nearest neighbor pairing $|\langle \hat c_{i+e_x}\hat c_{i}\rangle|$ as a
function of $T$ for various coupling strengths.  $|\langle \hat c_{i+e_x}\hat c_{i}\rangle|$: Pink $\times$'s for $g/t=0.7$, pink $\star$'s for $g/t=0.8$, and pink $\triangle$'s for $g/t=1.5$.  $\bar\rho_s$: Blue $\triangledown $'s for $g/t=0.7$, blue $\circ$'s for $g/t=0.8$, and blue $\diamond$'s for $g/t=1.5$.  
The numerical calculations are performed on a $27\times27$ lattice with one third filling. 
 Inset: The nearest neighbor pairing $|\langle \hat c_{i+e_x}\hat c_{i}\rangle|$ (green $\circ$'s) and the superfluid fractions
$\rho_{s,x}$ (pink $\times$'s) and $\rho_{s,y}$ (red $\star$'s) as a function of $g$ for  $T=0$. 
 }
\label{nsFig}
\end{figure}

In Fig.~\ref{nsFig}, we plot the superfluid fraction and the nearest neighbor order parameter as a function of $T$ for various coupling strengths. 
As usual for a 2D system, the superfluid fraction is discontinuous  at the critical temperature. Contrary to a translationally 
invariant system,  the superfluid fraction is less than 1, even for $T=0$~\cite{Paramekanti}.
In the inset, we plot the superfluid fraction and the nearest neighbor pairing as a function of coupling strength for $T=0$. We see that $\rho_{s,x}\neq\rho_{s,y}$, which  follows from the anisotropy of the $p$-wave paring. Note that the superfluid fraction 
behaves very differently from the pairing as a function of the coupling strength~\cite{Paananen}. 

We expect  correlation effects to decrease the transition temperatures of the ordered phases from what is predicted in the 
present paper. Even so,  we believe that our results  are qualitatively correct due to the long range nature of the interaction. This includes 
the scaling of $T_{\rm BKT}$, $T_c^{\rm st}$, and  $T_c^{\rm cb}$ for strong coupling. Our results therefore present a useful first analysis of the 
order phases of fermionic dipoles in a lattice at non-zero temperature.

\section{Trapped system}
The harmonic trapping potential  is always present  in atomic gas experiments. For $T=0$, this
leads to the co-existence of superfluid and density ordered phases forming ring and island structures~\cite{Gadsbolle}.
 We now investigate these effects at a non-zero temperature. 
 
\begin{figure}[th]
\includegraphics[clip=true,width=0.95\columnwidth]{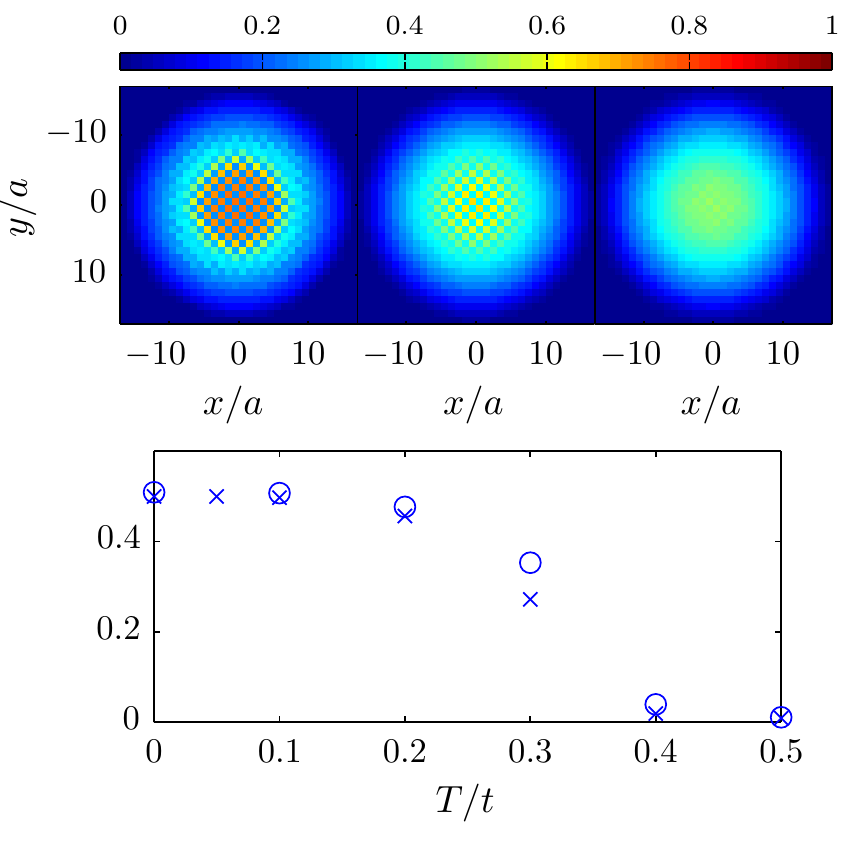}
\caption{(color on-line) Top: The density for $T/t=0$ (left), $T/t=0.3$ (middle), and  $T/t=0.4$ (right) for $g/t=1$,  $\tilde\omega=0.24$, 
 $\theta_P=0$, and $207-210$ dipoles trapped. 
Bottom: $\times$'s are the checkerboard order parameter $|\langle\hat n_{i}-\hat n_{i+e_y}\rangle|$ in the center of the trap as a function of $T$ and $\circ$'s are the checkerboard order parameter performed on the untrapped system at half-filling with the same parameters.  
 }\label{Traptheta0}
\end{figure}

Figure   \ref{Traptheta0} (top) shows the density and the checkerboard order parameter as a function of temperature
for the dipoles aligned perpendicularly to the lattice plane with  $(\theta_P,\phi_P)=(0,0)$. We have chosen $\tilde\omega=\omega a\sqrt{m/t}= 0.24$, $g/t=1$, 
and $\mu/t=4.23$ for the numerical calculations, giving $N=207-210$ dipoles trapped and an average filling fraction close to $1/2$ in the center of trap. 
For these parameters, there is a large  region in the center of the trap with checkerboard density order for $T=0$. 
 With increasing temperature, the radius of the checkerboard phase in the center shrinks and it melts completely for $T/t\simeq0.4$.  
In Fig.~\ref{Traptheta0} (bottom), we compare  the central value of the density order parameter with that of an un-trapped system at
 half-filling performed on a $30 \times 30$ lattice with the same interaction strength. We see that the critical temperature of the trapped system 
 is close to that of an untrapped system. This shows that the system essentially behaves according to the local density approximation. 

\begin{figure}
\includegraphics[clip=true,width=1\columnwidth]{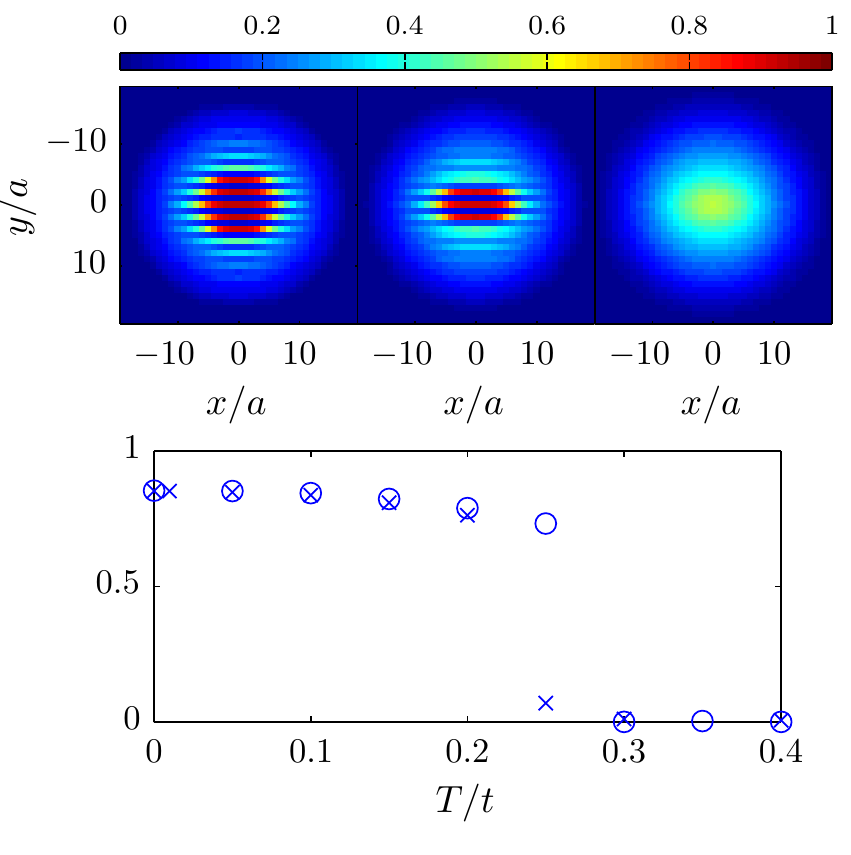}
\caption{(color on-line) Top: The density for $T/t=0$ (left), $T/t=0.1$ (middle), and  $T/t=0.3$ (right) for $g/t=1$,  $\tilde\omega=0.11$, 
 $(\theta_P,\phi_P)=(\pi/2,0)$, and $179-190$ dipoles trapped. 
Bottom: $\times$'s are the stripe order parameter $|\langle\hat n_{i}-\hat n_{i+e_y}\rangle|$ in the center of the trap as a function of $T$ and $\circ$'s are the stripe order parameter of an untrapped system at half filling with the same parameters. 
}
\label{Trapthetapi2}
\end{figure}

In Fig.~\ref{Trapthetapi2} (top), we plot the density and the stripe order parameter for the case where the 
dipoles are aligned along the $x$ axis with $(\theta_P,\phi_P)=(\pi/2,0)$. The coupling strength is $g/t=1$, $\tilde w=0.11$, and $\mu/t=-2$
giving $179-190$ dipoles trapped with the average filling $f=0.5$ in the center of trap. 
For this set of parameters, the center of the trap is in the striped phase for $T=0$. 
The stripe order disappears with increasing temperature. Interestingly, 
the melting is anisotropic in the sense that the stripe order disappears first in the 
$y$ direction. The stripe order is completely gone for $T/t\simeq 0.3$.  Again, we see from Fig.~\ref{Trapthetapi2} (bottom) that 
the density order in the center of the trap  agrees well with  that of an un-trapped system with the same parameters.

\begin{figure}
\includegraphics[clip=true,width=1\columnwidth]{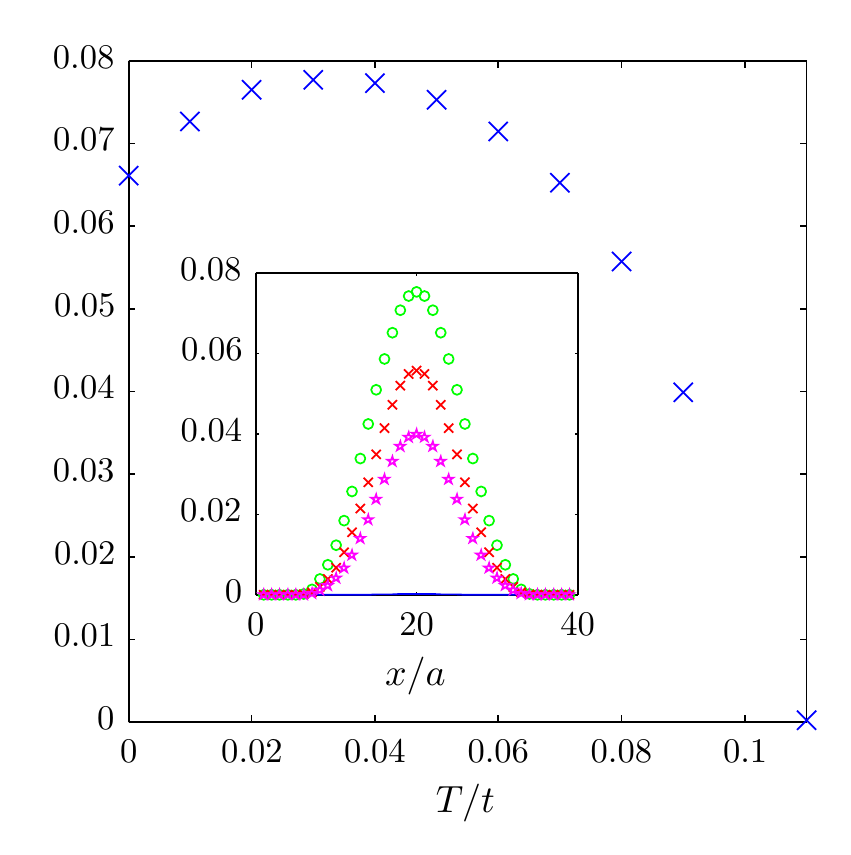}
\caption{(color on-line) The pairing order parameter $(|\langle \hat c_{i+e_x}\hat c_{i}\rangle|+|\langle \hat c_{i-e_x}\hat c_{i}\rangle|)/2$  in the centre 
for  $\mu=-1.72$ with $\theta_P=\pi/2$, $g/t=0.85$, and $\tilde\omega=0.11$ as a function of $T$. There are  $205-207$ dipoles trapped. 
The inset shows a diagonal cross section of the pairing order parameter. Green $\circ$'s are $T/t=0.05$, red $\times$'s are $T/t=0.08$, pink $\star$'s are $T/t=0.09$, and blue solid line is for $T/t=0.11$. }
\label{Trapthetapi/2Weak}
\end{figure}

Finally, we plot in Fig. \ref{Trapthetapi/2Weak} the pairing order parameter  as a function of temperature 
for  $\mu/t=-1.72$ with $(\theta_P,\phi_P)=(\pi/2,0)$, $g/t=0.85$, $\tilde\omega=0.11$, and $205-207$ dipoles trapped.  
Since the coupling is weak, the system is superfluid for $T=0$ and there is no stripe order. As expected, 
the pairing decreases with increasing $T$  and it disappears for $T/t\simeq0.11$. The critical temperature is calculated using mean-field 
theory. We expect corrections to mean-field theory to be small  since the critical temperature  is so small. The pairing increases slightly with increasing $T$
at low temperature. 
This is because we for simplicity  keep the chemical potential fixed in the numerical calculations leading to an increased density with increasing $T$. 
A number conserving calculation would yield a monotonically decreasing pairing with increasing $T$.

These results illustrate that even in the presence of a trap, one can observe the superfluid 
and density ordered phases predicted for the infinite lattice systems, provided the system is large enough. In particular, the transition temperature is 
determined by the parameters in the center of the trap, and the results for a system with no trap can be used.

\section{Conclusion} 
In conclusion, we  examined the density ordered and superfluid phases of fermionic dipoles  in a square 2D lattice. We determined 
the  critical temperature of the density ordered phases and demonstrated that it is  proportional to the interaction strength for strong coupling. 
We calculated the superfluid fraction and showed that the critical temperature of the superfluid phase is proportional to the hopping matrix element
for strong coupling. 
Finally, we analyzed the effects of the harmonic trapping potential showing that for systems of a realistic size, the density 
ordered and superfluid phases exist with critical temperatures  close to those obtained from a local density 
approximation. 

A.-L. G. is grateful to N.\ Nygaard for valuable discussions concerning the superfluid density and to S.\ Gammelmark for Fig. \ref{setup}.

\end{document}